\documentclass[twocolumn,showpacs,amsmath,amstex,amssymb,mathfonts,prl]{revtex4-1}
\usepackage{graphicx,bm,units}

\usepackage{color}

\begin{document}

\newcommand{\vk}{{\bf k}}
\def\ns{^{\vphantom{*}}}
\def\ket#1{{|  #1 \rangle}}
\def\bra#1{{\langle #1|}}
\def\braket#1#2{{\langle #1  |   #2 \rangle}}
\def\expect#1#2#3{{\langle #1 |   #2  |  #3 \rangle}}
\def\cH{{\cal H}}
\def\half{\frac{1}{2}}
\def\sut{\textsf{SU}(2)}
\def\suto{\textsf{SU}(2)\ns_1}
\def\kF{\ket{\,{\rm F}\,}}

\title{Topological Entanglement in Abelian and non-Abelian Excitation Eigenstates}

\author{Z. Papi\'c$^{1,2,3}$, B. A. Bernevig$^4$ and N. Regnault$^3$}
\affiliation{$^1$ Institute of Physics, University of Belgrade, Pregrevica 118, 11 000 Belgrade, Serbia}
\affiliation{$^2$ Laboratoire de Physique des Solides, Univ. Paris-Sud, CNRS, UMR 8502, 91405 Orsay, France} 
\affiliation{$^3$ Laboratoire Pierre Aigrain, ENS and CNRS, 24 rue Lhomond, F-75005 Paris, France}
\affiliation{$^4$ Department of Physics, Princeton University, Princeton, NJ 08544}

\begin{abstract} 
Entanglement in topological phases of matter has so far been investigated through the perspective of their ground-state wavefunctions.
In contrast, we demonstrate that the \emph{excitations} of Fractional Quantum Hall (FQH) systems also contain information to identify the system's topological order. Entanglement spectrum of the FQH quasihole (qh) excitations is shown to differentiate between the Conformal Field Theory (CFT) sectors, based on the relative position of the qh with respect to the entanglement cut. For Read-Rezayi model states, as well as Coulomb interaction eigenstates, the counting of the qh entanglement levels in the thermodynamic limit matches exactly the CFT counting, and 
sector changes occur as non-Abelian quasiholes successively cross the entanglement cut.
\end{abstract}


\date{\today}

\maketitle

Topologically ordered systems are not characterized by local order parameters; non-local concepts, such as quantum entanglement \cite{entanglement}, have been extensively used  in recent years to describe such phases of matter. The favorite method of analyzing the entanglement -- entanglement entropy (or its topological part for gapped systems \cite{kitaev_levin}) -- does not result in a unique characterization of the system: different states of matter can  have identical entanglement entropy. Complicated topological phases, such as FQH states, are fully described by a multitude of universal parameters, notably braiding matrices \cite{difrancesco}, which are related to the properties of the FQH \emph{excitations} under adiabatic exchanges in space-time. In finite systems, the braiding matrices are impossible to obtain and the question arises whether the universal properties of a topologically ordered state are obtainable via the entanglement of its \emph{excitations}. Although scarcely addressed in the existing literature \cite{huse2010, fradkin}, the question is pertinent also in view of the phases of matter that can only be distinguished by their excitation spectra \cite{freedman}.

\begin{figure}[htb]
\centering
\includegraphics[width=0.7\linewidth]{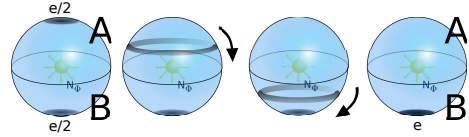}
\caption{FQH sphere with a monopole of $N_\Phi$ magnetic flux quanta in the center, partitioned into hemispheres $A$ and $B$, and containing the bosonic Moore-Read state with two qhs. We start with two separated qhs (left) and drag one qh from the north 
to the south hemisphere (middle). The moving qh is azimuthally delocalized. We end up with a qh twice the charge at 
the south pole (right).} \label{sphere}
\end{figure}

Recently, it was proposed \cite{li_haldane} that the \emph{entanglement spectrum} (ES), {\it i.e.\/} the (negative logarithm of the) full set of eigenvalues of the reduced density matrix $\rho\ns_A$ is a rich source of information on the topological order in FQH ground states. Reduced density matrix $\rho\ns_A$ of the subsystem $A$ of a pure FQH state $|\psi\rangle$ on the sphere \cite{haldane_sphere} (Fig. \ref{sphere}) is given by the usual trace $\rho\ns_A = {\rm Tr}_B  |\psi\rangle \langle \psi |$ over the complementary subsystem $B$. The levels of $\rho\ns_A$ can be classified according to the number of particles $N_A$ and orbitals $l_A$ in $A$, as well as the $z$-axis projection of the angular momentum, $L_z^A$. The multiplicities and relative energy spectrum of $\rho\ns_A$ matches that of the edge modes \cite{li_haldane,regnault2009, thomale2010,lauchli}. For ground states of Coulomb Hamiltonians in the same universality class with a FQH model state, the ES typically displays a branch  of low-lying (high probability) levels, very similar to those of the model state, accompanied by spurious levels at high entanglement energy (low probability). The gap between the low and high levels, properly defined by taking the ``conformal limit" \cite{thomale2010}, was conjectured and numerically substantiated to remain constant upon increasing the system size. 

\begin{figure*}[t]
  \begin{minipage}[l]{\linewidth}
    \includegraphics[width=\linewidth]{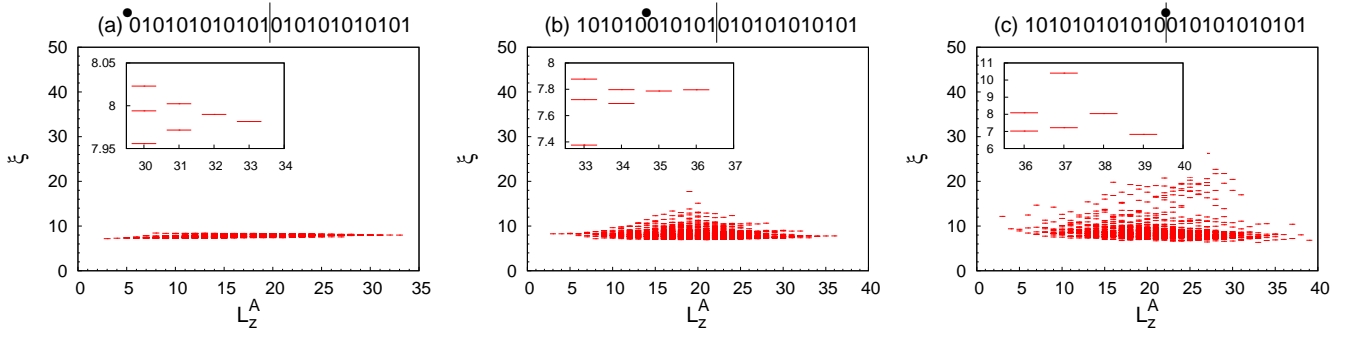}
  \end{minipage}
\caption{(Color online) Conformal-limit ES of the Laughlin model state of $N=12$ bosons with a single qh, localized at one of the poles (a),  the equator (c) and in between (b). The location of the qh is given by the dot above each root partition.}
\label{laughlin_onehalf_m_0_6_12}
\vspace{-0pt}
\end{figure*}
In this Letter we show that the ES of FQH \emph{excitation} states contains information to identify the universal properties of topological phases of matter. We consider model wavefunctions, such as Laughlin \cite{laughlin}, Moore-Read \cite{mr} and Read-Rezayi \cite{rr}, whose ground- and excited states with localized qhs can be expressed as Jack polynomials \cite{jack, benoit}. Furthermore, we consider the eigenstates of Coulomb interaction potential with the impurities that pin the qhs at a circle of latitude \cite{deltaimpurity, prodan_haldane}. The ES of a given FQH excitation is monitored as the qhs are moved across the cut (Fig. \ref{sphere}). This reveals that the ES of the excitations can probe different CFT fermion number sectors, that it gives the correct counting of the edge states in the thermodynamic limit, and that it is extremely sensitive (within a single magnetic length)  to whether non-Abelian qhs are on the same or opposite sides of the entanglement cut. The latter property can be taken as a simple manifestation of the non-Abelian nature of the phase. These findings are corroborated in studies of the realistic Coulomb interaction eigenstates. We also analyze the behavior of the quasielectron excitations, as well as the quasihole and quasielectron fluctuations when close to the entanglement cut.  

\begin{figure}[thb]
\centering
\includegraphics[scale=0.5]{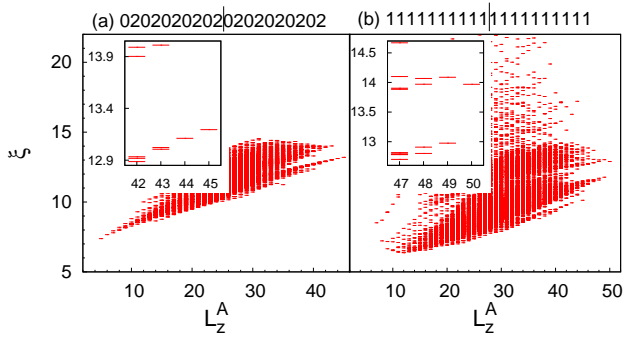}
\caption{(Color online) Conformal-limit ES of the Moore-Read model state for $N=20$ bosons with a unit flux added.  (a) Abelian vortex $0202\ldots 02$. (b) Two fractionalized $e/2$ non-Abelian qhs.}
\label{mr_one_m_0_10}
\end{figure}
\begin{figure}[thb]
\centering
\includegraphics[scale=0.5]{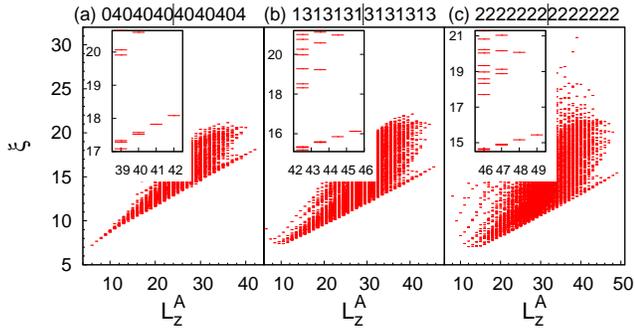}
\caption{(Color online) Conformal-limit ES of the $\mathbb{Z}_4$ Read-Rezayi state of $N=28$ bosons and a unit flux added. Abelian vortex (a) is fractionalized and the non-Abelian $e/4$ qhs are moved to the opposite pole one at a time ((b),(c)), causing the sector change.} \label{z4}
\end{figure}
\begin{figure*}[t]
  \begin{minipage}[l]{\linewidth}
    \includegraphics[width=\linewidth]{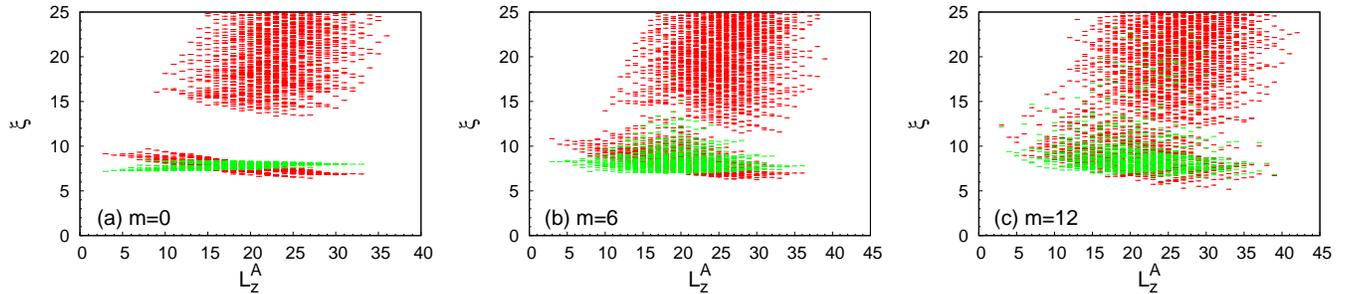}
  \end{minipage}
\caption{(Color online) Conformal-limit ES of the Coulomb $\nu=1/2$ state of $N=12$ bosons with a single qh, localized using delta impurity potential, at one of the poles $m=0$ (a),  the equator $m=12$ (c) and in between $m=6$ (b). The cut is defined by $N_A=6$ and $l_A=12$.}
\label{coulomb_onehalf}
\vspace{-0pt}
\end{figure*}
The orbital of an electron confined to the lowest Landau level (LLL)  moving on the Haldane sphere \cite{haldane_sphere} (Fig.\ref{sphere}) can be written as $\phi_m (z)= \mathcal{N}_m z^m $, where $z=x+iy$ is the complex 2D electron coordinate, the quantum number $m$ is the $L_z$ eigenvalue, and $\mathcal{N}_m$ is a geometrical normalization factor. ``Conformal limit" \cite{thomale2010} is defined as the geometry where the normalization factors $\mathcal{N}_m$ are all equal to 1. General $N$-electron LLL states are analytic polynomials $\Psi_F(z_1,\ldots,z_N) = \prod_{i<j} (z_i-z_j) \times \Psi_B$, which can be factorized into a Vandermonde determinant and a bosonic wave function $\Psi_B$. We may therefore focus on the systems of charged bosons in the LLL. Arbitrary $\Psi_B$ is expandable  in terms of symmetric monomials indexed by a partition $\lambda$ represented by occupation numbers $n(\lambda) = \{n_m(\lambda), m=0,1,..\}$ of the orbitals $\phi_m$. Certain FQH wavefunctions are, however, defined by a single \emph{root} partition -- all the remaining partitions  in the expansion of $\Psi_B$ are derived from it via ``squeezing" operations \cite{jack}. This includes the Read-Rezayi \cite{rr} $\mathbb{Z}_k$ series of trial states for bosons at filling factors $\nu\equiv N/N_\Phi=k/2$ which can be identified with a family of Jack polynomials (Jacks) $J_{\lambda_k}^\alpha$, parameterized by $\alpha=-(k+1)$ and indexed by a partition $n(\lambda_k) \equiv (k0k0k\ldots)$ i.e. $\Psi_{\rm RR}^k  \propto J_{k0k0\ldots }^{-(k+1)}\left( \{ z_i \} \right)$.
Apart from FQH ground states, the Jacks also yield wavefunctions of the qh \cite{jack} and quasielectron \cite{jack_qe} excitations, created by varying the magnetic flux through the sphere. 
If the flux is increased by one unit, $k$ non-Abelian qhs of charge $e/k$ each are nucleated and pinned at locations $w_1,\ldots, w_k$. Let us consider such an arrangement of the qhs where the first $n_1$ qhs are fixed at one pole of the sphere, $n_2=k-1-n_1$ are at the opposite pole and the remaining mobile qh is at $w_k$ in between the poles. The wavefunction for $\mathbb{Z}_k$ states with a qh at $w_k$ close to {\it e.g.} the north pole is given by a single Jack $J_{abab\ldots ab}^{-(k+1)} \left(  \{ z_i \} \right)$ with the root partition $abab\ldots ab$, where $a=(k-1-\Delta n)/2$, $b=k-a$ and $\Delta n = n_2-n_1$ \cite{benoit}. As the qh at $w_k$ is moved from the north to the south pole, the wavefunction mixes in several other Jacks \cite{jack, benoit}, and we can track its progression. When the qh reaches the south pole, the wavefunction is again a single Jack, $J_{a+1b-1a+1b-1\ldots a+1 b-1}^{-(k+1)} \left( \{ z_i \} \right)$. 

Jack qh wavefunction for $k=1$ describes a single Abelian qh, localized at one of the orbitals of the sphere. In Fig. \ref{laughlin_onehalf_m_0_6_12} we show the numerically calculated ES in the conformal limit for the Laughlin state of $N=12$ bosons and a single localized qh. The cut is fixed such that the subsystem $A$ contains $l_A=12$ orbitals and $N_A=6$ particles. 
Above each plot in Figs. \ref{laughlin_onehalf_m_0_6_12}-\ref{z4}, we give the root configuration used to generate the wavefunction and draw the cut in orbital space (vertical line), showing the high-probability ES levels in the inset. 
When the qh is at one of the poles (Fig. \ref{laughlin_onehalf_m_0_6_12}(a)), the level counting matches that of the $U(1)$ chiral boson CFT. As we move the qh towards the equator (Fig. \ref{laughlin_onehalf_m_0_6_12}(b)), the levels spread upwards but the CFT counting remains unaltered.  The spreading reflects the quantum-mechanical oscillation of the qh around the cut but is expected to be confined within a magnetic length around the equator; away from the equator, the counting of the qh entanglement levels remains a faithful  representation of the FQH state. 
The  CFT counting is lost \emph{only} when the qh is situated \emph{exactly} on the entanglement cut, which effectively splits the qh in two (Fig. \ref{laughlin_onehalf_m_0_6_12}(c)). Using Jacks we obtained the ES of larger systems than those attainable by exact diagonalization and $N=12$ is presented only to facilitate comparison with the Coulomb case below. 

For $k=2$ Moore-Read state, two non-Abelian qhs are formed when a unit flux is added. When both qhs are at the same pole, the wavefunction is represented by the root  $0202\ldots 02$.  When a single $e/2$ qh is fixed at one pole and the other is moved to the opposite pole, the Jack root is $1111\ldots 11$. ES for these two cases is shown in Fig. \ref{mr_one_m_0_10} and demonstrates that a \emph{sector change} has taken place as the non-Abelian qh is moved across the cut: the counting $1,1,3\ldots$ has changed into $1,2,4\ldots$ The two different countings represent the two topological sectors of the theory, the even and odd fermion-number sectors respectively. Sector change occurs immediately as the qh crosses the entanglement cut.  The thermodynamic-limit counting of the two \emph{excitation} wavefunctions is in one to one correspondence to  that  of the excitations above the ground state for an even ($20202000\ldots$) and odd ($20201000\ldots$) number of particles. 

An interesting question is what happens when  more than two topological sectors are present in the theory. 
The simplest example is the $\mathbb{Z}_4$ Read-Rezayi state where the three sectors' counting is given by the excitations above the ground states with $N=0,1,2 \;(\text{mod} \; 4) $ number of particles:  $4040404000\ldots$, the $\sigma_1$ sector $4040404010000\ldots$ (equivalent to the $\sigma'_1$ sector $40404040300000\ldots$) and the $\sigma_2$ sector $4040404020000\ldots$ ($\sigma_2$ is its own conjugate). In Fig. \ref{z4} we show the ES for the $\mathbb{Z}_4$ state of $N=28$ bosons with a single flux added. Starting from an Abelian vortex localized at one of the poles $0404\ldots 04$ (Fig. \ref{z4}(a)) with the counting $1,1,3,5,\ldots$, we transfer a single non-Abelian qh to the opposite pole. If we do this once (Fig. \ref{z4}(b)), we obtain the state with the root $1313\ldots 13$ and the counting $1,2,5,8\ldots$; doing it once again results in a root $2222\ldots 22$ and the counting $1,2,6,9,\ldots$ (Fig. \ref{z4}(c)). 
 
Our main findings for the model states -- the counting of the qh edge modes and the sector change upon crossing the cut -- carry over  to the realistic systems with LLL-projected Coulomb interaction. The problem of qh localization was solved generally in Ref. \onlinecite{prodan_haldane}. For the Laughlin filling $\nu=1/2$,  we simulate the probe by a delta-function impurity potential \cite{deltaimpurity} of weight $0.005/\sqrt{N_\Phi/2}$, localized on the orbital $m$  (Fig. \ref{sphere}). This potential is weak enough not to cause any level crossing in the spectrum, but only splits the degenerate multiplets of states. In Fig. \ref{coulomb_onehalf} we show the ES of the Coulomb ground state in the presence of a localized qh by superimposing it on the model state from the Fig. \ref{laughlin_onehalf_m_0_6_12}. A finite entanglement gap separates the CFT branch from the generic Coulomb continuum of higher entanglement-energy states. CFT branch, however, displays the same counting as the model state. The Coulomb continuum largely remains fixed as we move the qh towards the equator, indicating  that the scatter of energy levels in Fig. \ref{coulomb_onehalf} is essentially similar to the qh oscillation across the cut and therefore due to the conformal levels spreading in the upward direction. The Coulomb qh therefore has the same behavior as the model one, at least  sufficiently far from the cut.

\begin{figure}[htb]
\centering
\includegraphics[scale=0.5]{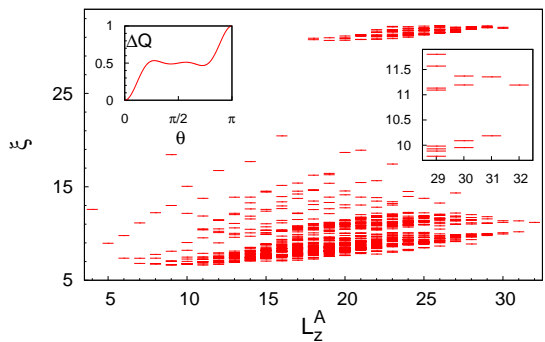}
\caption{(Color online) Conformal-limit ES of the Coulomb $\nu=1$ state for $N=16$ bosons and a unit flux added. Delta impurity is located at each pole to pin the non-Abelian qh, as seen in the $e/2$ step in the plot of the excess charge (left inset). The cut is defined by $N_A=8$ and $l_A=8$. Right inset shows the zoom on the lowest ES levels.}      \label{coulomb_one}
\end{figure}
We next study the $\nu=1$ Coulomb state, expected to be in the universality class of Moore-Read.  
Abelian vortex gives identical counting to the one derived from the root partition $0202\ldots 02$ in Fig. \ref{mr_one_m_0_10} \cite{regnaultjain}. To separate the two non-Abelian qhs, one on each pole, we use the method of Ref. \onlinecite{prodan_haldane} restricting the Hilbert space to the $L_z=0$ sector of the Moore-Read zero-modes. Within this subspace, we use a combination of two delta impurities, one on each pole, to trap the non-Abelian qhs \cite{regnaultjain}. To ensure that the qhs are indeed separated, we calculate the excess charge $\Delta Q (\theta)$ \cite{regnaultjain}. Note that for the fermionic Moore-Read state, a two-body potential is necessary in order to pin the qhs \cite{prodan_haldane}. In Fig. \ref{coulomb_one} we recognize the same counting as the one that can be derived from the root configuration $1111\ldots 11$, which confirms that the sector change has occurred.

\begin{figure}[htb]
\centering
\includegraphics[scale=0.5]{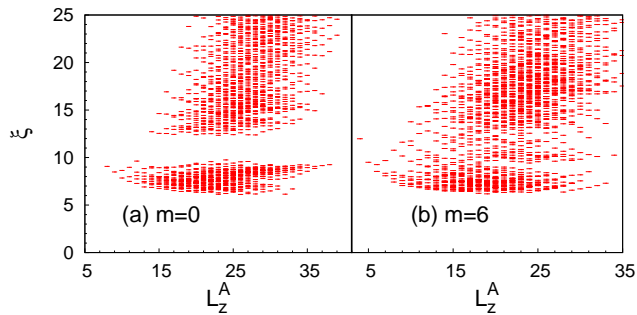}
\caption{(Color online) Conformal-limit ES of the Laughlin model state of $N=12$ bosons on the sphere with a unit flux removed. The quasielectron is locked to a delta impurity at one the poles ($m=0$) and displaced half way toward the equator ($m=6$).}
\label{quasielectron_laughlin_onehalf_m_0_6_12}
\end{figure}
Finally, we have also analyzed the ES of the Laughlin model \emph{quasielectron} states, Fig. \ref{quasielectron_laughlin_onehalf_m_0_6_12}, obtained as ground states of the pseudopotential Hamiltonian at one flux removed compared to the ground state and with a delta impurity. Unlike the qh, quasielectron states show a finite entanglement gap. This is expected as they are not unique and densest zero modes of any local Hamiltonian \cite{jack}. Although the counting of the levels is again a faithful representation of the edge spectrum, the effect of oscillation across the cut is much more pronounced due to their larger size \cite{nuebler} compared to the corresponding qh excitation. This leads to a rapid closing of the entanglement gap as the quasielectron approaches the equator.

This work was supported by the Agence Nationale de la Recherche under Grant No. ANR-JCJC-0003-01. 
ZP acknowledges support by the Serbian Ministry of Science under Grants No. 141035 and No. ON171017,
and by grants from R\'egion Ile-de-France. BAB was supported by Princeton Startup Funds, Alfred P. Sloan
Foundation, NSF DMR-095242, and NSF China 11050110420, and MRSEC grant at Princeton
University, NSF DMR-0819860. BAB thanks the Ecole Normale Sup\'erieure, Paris for generous hosting.

\end{document}